\documentstyle[preprint,aps,epsf]{revtex}
\begin{document}
\draft

\author{
    E.~M.~Aitala,$^9$
       S.~Amato,$^1$
    J.~C.~Anjos,$^1$
    J.~A.~Appel,$^5$
       D.~Ashery,$^{15}$
       S.~Banerjee,$^5$
       I.~Bediaga,$^1$
       G.~Blaylock,$^8$
    S.~B.~Bracker,$^{16}$
    P.~R.~Burchat,$^{14}$
    R.~A.~Burnstein,$^6$
       T.~Carter,$^5$
 H.~S.~Carvalho,$^{1}$
  N.~K.~Copty,$^{13}$
    L.~M.~Cremaldi,$^9$
 C.~Darling,$^{19}$
       K.~Denisenko,$^5$
       A.~Fernandez,$^{12}$
       P.~Gagnon,$^2$
       K.~Gounder,$^9$
     A.~M.~Halling,$^5$
       G.~Herrera,$^4$
 G.~Hurvits,$^{15}$
       C.~James,$^5$
    P.~A.~Kasper,$^6$
       S.~Kwan,$^5$
    D.~C.~Langs,$^{11}$
       J.~Leslie,$^2$
       B.~Lundberg,$^5$
       S.~MayTal-Beck,$^{15}$
       B.~Meadows,$^3$
 J.~R.~T.~de~Mello~Neto,$^1$
    R.~H.~Milburn,$^{17}$
 J.~M.~de~Miranda,$^1$
       A.~Napier,$^{17}$
       A.~Nguyen,$^7$
  A.~B.~d'Oliveira,$^{3,12}$
       K.~O'Shaughnessy,$^2$
    K.~C.~Peng,$^6$
    L.~P.~Perera,$^3$
    M.~V.~Purohit,$^{13}$
       B.~Quinn,$^9$
       S.~Radeztsky,$^{18}$
       A.~Rafatian,$^9$
    N.~W.~Reay,$^7$
    J.~J.~Reidy,$^9$
    A.~C.~dos Reis,$^1$
    H.~A.~Rubin,$^6$
 A.~K.~S.~Santha,$^3$
 A.~F.~S.~Santoro,$^1$
       A.~J.~Schwartz,$^{11}$
       M.~Sheaff,$^{18}$
    R.~A.~Sidwell,$^7$
    A.~J.~Slaughter,$^{19}$
    M.~D.~Sokoloff,$^3$
       N.~R.~Stanton,$^7$
       K.~Stenson,$^{18}$
    D.~J.~Summers,$^9$
 S.~Takach,$^{19}$
       K.~Thorne,$^5$
    A.~K.~Tripathi,$^{10}$
       S.~Watanabe,$^{18}$
 R.~Weiss-Babai,$^{15}$
       J.~Wiener,$^{11}$
       N.~Witchey,$^7$
       E.~Wolin,$^{19}$
       D.~Yi,$^9$
       S. Yoshida,$^{7}$                         
       R.~Zaliznyak,$^{14}$
       and
       C.~Zhang$^7$ \\
\begin{center} (Fermilab E791 Collaboration) \end{center}
}

\address{
$^1$ Centro Brasileiro de Pesquisas F{\'i}sicas, Rio de Janeiro, Brazil\\
$^2$ University of California, Santa Cruz, California 95064\\
$^3$ University of Cincinnati, Cincinnati, Ohio 45221\\
$^4$ CINVESTAV, Mexico\\
$^5$ Fermilab, Batavia, Illinois 60510\\
$^6$ Illinois Institute of Technology, Chicago, Illinois 60616\\
$^7$ Kansas State University, Manhattan, Kansas 66506\\
$^8$ University of Massachusetts, Amherst, Massachusetts 01003\\
$^9$ University of Mississippi, University, Mississippi 38677\\
$^{10}$ The Ohio State University, Columbus, Ohio 43210\\
$^{11}$ Princeton University, Princeton, New Jersey 08544\\
$^{12}$ Universidad Autonoma de Puebla, Mexico\\
$^{13}$ University of South Carolina, Columbia, South Carolina 29208\\
$^{14}$ Stanford University, Stanford, California 94305\\
$^{15}$ Tel Aviv University, Tel Aviv, Israel\\
$^{16}$ 317 Belsize Drive, Toronto, Canada\\
$^{17}$ Tufts University, Medford, Massachusetts 02155\\
$^{18}$ University of Wisconsin, Madison, Wisconsin 53706\\
$^{19}$ Yale University, New Haven, Connecticut 06511
}
\title{ 
       Measurement of the branching ratio 
      ${{\cal B}(D^+\rightarrow\rho^0\ell^+\nu_\ell)}/
       {{\cal B}(D^+\rightarrow\overline{K}^{*0}\ell^+\nu_\ell)}$
      }

\date{\today}

\maketitle

\begin{abstract}
      We report a measurement of the branching ratio  
      ${{\cal B}(D^+\rightarrow\rho^0\ell^+\nu_\ell)}$/ \newline
      ${{\cal B}(D^+\rightarrow\overline{K}^{*0}\ell^+\nu_\ell)}$ 
      from the Fermilab charm hadroproduction experiment E791. 
      Based on signals of $49\pm17$ events in the
      $D^+\rightarrow\rho^0 e^+\nu_e$ mode and $54\pm18$ events in the
      $D^+\rightarrow\rho^0 \mu^+\nu_\mu$ mode, we measure  
\begin{eqnarray*}      
       {{\cal B}(D^+\rightarrow\rho^0 e^+\nu_e)}/
       {{\cal B}(D^+\rightarrow\overline{K}^{*0}e^+\nu_e)} & = &
      0.045\pm0.014\pm0.009 \mbox{, and} \\
       {{\cal B}(D^+\rightarrow\rho^0 \mu^+\nu_\mu)}/
       {{\cal B}(D^+\rightarrow\overline{K}^{*0}\mu^+\nu_\mu)} & = &
      0.051\pm0.015\pm0.009.
\end{eqnarray*} 
      Combining the results from both the 
      electronic and muonic modes, we obtain 
      $${{\cal B}(D^+\rightarrow\rho^0\ell^+\nu_\ell)}/
       {{\cal B}(D^+\rightarrow\overline{K}^{*0}\ell^+\nu_\ell)}= 
      0.047\pm0.013.$$ 
This result is compared to theoretical predictions.
\end{abstract}

\pacs{13.20.-v, 13.20.Fc, 13.30.Ce}
%
%
\newcommand{\kstln}{\mbox{$D^+\rightarrow\overline{K}^{*0}\ell^+\nu_\ell$}}
\newcommand{\rholn}{\mbox{$D^+\rightarrow\rho^0\ell^+\nu_\ell$}}
\newcommand{\ksten}{\mbox{$D^+\rightarrow\overline{K}^{*0}e^+\nu_e$}}
\newcommand{\rhoen}{\mbox{$D^+\rightarrow\rho^0 e^+\nu_e$}}
\newcommand{\kstmn}{\mbox{$D^+\rightarrow\overline{K}^{*0}\mu^+\nu_\mu$}}
\newcommand{\rhomn}{\mbox{$D^+\rightarrow\rho^0 \mu^+\nu_\mu$}}
\newcommand{\philn}{\mbox{$D_s^+\rightarrow\phi\ell^+\nu_\ell$}}
\newcommand{\Fetapln}{\mbox{$D_s^+\rightarrow\eta'\ell^+\nu_\ell$}}
\newcommand{\Detapln}{\mbox{$D^+\rightarrow\eta'\ell^+\nu_\ell$}}
\newcommand{\Fetaln}{\mbox{$D_s^+\rightarrow\eta\ell^+\nu_\ell$}}
\newcommand{\Detaln}{\mbox{$D^+\rightarrow\eta\ell^+\nu_\ell$}}
\newcommand{\omegaln}{\mbox{$D^+\rightarrow\omega\ell^+\nu_\ell$}}
\newcommand{\dpkpp}{\mbox{$D^+\rightarrow K^- \pi^+\pi^+$}}
\newcommand{\dpppp}{\mbox{$D^+\rightarrow \pi^-\pi^+\pi^+$}}

\narrowtext

Semileptonic charm decays are useful in probing the dynamics of 
hadronic currents since the Cabibbo-Kobayashi-Maskawa matrix elements
for the charm sector are 
well-known from unitarity constraints. Form factors 
for Cabibbo-suppressed (CS) $c\rightarrow d$ semileptonic decays can
be related via Heavy Quark Effective Theory (HQET) 
to those for $b\rightarrow u$ semileptonic decays
at the same four-velocity transfer\cite{HQET}. 
Since knowledge of the form factors in
 $b\rightarrow u$ transitions is vital for extracting $V_{ub}$ 
from $b\rightarrow u$ semileptonic decays in a model-independent way, 
study of $c\rightarrow d$ semileptonic decays can improve our knowledge
of $V_{ub}$.
Although considerable progress has been made in studying
CS semileptonic charm decays to pseudoscalar mesons\cite{piln}, 
the only previous result on CS semileptonic charm decay to a vector
meson is based on four \rhomn\ events\cite{e653-rho}. 
In this Letter, we report a new measurement from the Fermilab hadroproduction
experiment E791 of ${{\cal B}(\rholn)}/{{\cal B}(\kstln)}$
based on more than 100 \rholn\ decays in the combined electronic and muonic
modes.

The E791 experiment\cite{e791-det} recorded $2\times10^{10}$ events   
from 500 GeV/$c$ $\pi^{-}$ interactions in five thin targets 
(one platinum, four diamond) separated by gaps of 1.34 to 1.39 cm.
Precision tracking and vertexing information was
provided by 23 silicon microstrip detectors (6 upstream and 17
downstream of the targets) and 35 drift chamber
planes. 
Momentum was measured with two dipole magnets. 
Two segmented threshold \v{C}erenkov counters provided $\pi/K$ separation in
the $6-60$\,GeV/$c$ momentum range\cite{Cerenkov}. 

Candidates for \rholn, $\rho^0\rightarrow\pi^+\pi^-$ and 
\kstln, $\overline{K}^{*0}\rightarrow K^-\pi^+$ decays
(charge-conjugate states are implied throughout this Letter) 
are selected by requiring a three-prong decay vertex of 
charge $\pm1$ with one of the decay particles being identified as a
lepton. 
A segmented lead and liquid-scintillator calorimeter\cite{calor} 
is used to identify the 
electrons, based on energy deposition and transverse shower shape. The
probability that a $\pi$ ($K$) is misidentified as an electron is about
 $0.8\%$ ($0.5\%$). 
Muon identification is provided by
two planes of scintillation counters oriented horizontally and vertically, 
located behind
shielding with a thickness equivalent to 2.5 meters of steel (15
interaction lengths). All the muon candidates are required to have 
momentum greater than 12 GeV/$c$ to reduce background from decays in
flight. The
probability that a $\pi$ ($K$) is misidentified as a $\mu$ is about $1.6\%$
($2.4\%$). 

Once the lepton is identified, the other two tracks in the vertex
($h_1,h_2$) are assigned hadron masses.
We define the right-sign (RS) sample as vertices in which the lepton
and $D^+$ candidate have the same charge; $h_1$ and $h_2$ are then
oppositely-charged. 
For \kstln~candidates the hadron with odd charge is assigned a kaon mass,
while for \rholn~candidates $h_1$ and $h_2$ are both assigned pion
masses.
The RS sample contains both signal and backgrounds from
reconstruction errors and other charm decay channels.
The wrong-sign (WS) sample, in which the lepton and 
$D^+$ candidate have opposite
charge and $h_1,h_2$ have like charge, 
provides an estimate of the shape of the background under the 
 $\rho^0$ ($K^{*0}$) peak in the RS sample. 
Both kaon-assignment hypotheses are kept for WS \kstln~candidates.
 
Due to the undetected
neutrino in the $D^+$ decay, there are two solutions 
for the $D^+$ momentum. We choose the lower-momentum solution 
since Monte Carlo studies show that, for the $D^+$ 
three-prong semileptonic decays, this solution has a slightly larger
probability to be correct and  
offers somewhat better $D^+$ momentum resolution. 

To minimize systematic uncertainties, 
most selection criteria for \rholn\ are identical to those for 
\kstln; exceptions are discussed below. 
In addition, all criteria except those for lepton identification 
are identical for electronic and muonic decays. 
The common criteria are the following. A decay vertex must be 
separated from the production vertex by at least 20$\sigma_l$, where
$\sigma_l$ is the error on the measured separation. The decay vertex is
required to be at least 5$\sigma_m$ outside the nearest
solid material, where $\sigma_m$ is the error on the measured distance. 
The proper decay time for the $D^+$ candidate is required to be
less than 5\,ps. 
The hadron candidates in the decay are required to have momenta
greater than 6\,GeV/$c$. 
The minimum kinematically-allowed parent mass for 
the candidate $D^+\rightarrow h_1 h_2 \ell \nu_\ell$
decay, $M_{min}(h_1h_2\ell\nu_\ell)=p_T+\sqrt{p^2_T+ M_{vis}^2}$, is
required to lie between 1.6 and 2.0\,GeV/$c^2$, where
$p_T$ is the transverse momentum of $h_1h_2\ell$ with respect to
the $D^+$ flight direction and $M_{vis}$ is the invariant mass
of $h_1h_2\ell$.  
When masses are correctly assigned, the $M_{min}$ distribution has a
cusp at the $D^+$ mass. 
This distribution is broadened and shifted to lower mass if there are      
additional neutral hadrons in the final state.
The potential feedthroughs from hadronic decays such as \dpkpp\ and
\dpppp\ are removed explicitly by excluding candidates
with either $K\pi\pi$ or $\pi\pi\pi$ invariant mass within
$20$\,MeV/$c^2$ of the $D^+$ mass. 
Feedthrough from the Cabibbo-favored (CF) decay \philn\ followed by 
 $\phi\rightarrow K^+K^-$ is eliminated by excluding the region
 between 1.01 and 1.03\,GeV/$c^2$ in the $K^+K^-$ invariant mass. 

In the rarer \rholn\ mode, additional selection criteria are required 
to reduce non-charm background and to eliminate feedthrough
from the CF \kstln\ mode, which has a rate 20 times larger. 
The following criteria, applied only to \rholn, further reduce
non-charm background.
The maximal missing mass squared,
$M^2_{miss}=M^2_D+M^2_{vis}-2M_D\sqrt{M^2_{vis}+p^2_T}$, is required
to be in the range $-0.10$ to 0.15\,(GeV/$c^2$)$^2$. 
The scalar sum of the transverse momenta of the daughter
tracks with respect to the $D^+$ flight direction is required 
to be greater than 1.0\,GeV/$c$. 
Although these
quantities are partially correlated with the minimum parent mass, 
they do provide additional discriminating power.

When a $K$ from the CF mode \kstln\ is misidentified as a $\pi$, 
the reflected di-pion invariant mass is similar in position and shape
to the $\rho^0$ resonance.  
It is thus imperative to reduce contamination from \kstln\ to a level
well below the signal.
This is achieved with three selection criteria applied to candidate
\rholn\ decays, but not to the normalizing mode. 
1) The minimum parent mass computed 
for a $K\pi\ell\nu_\ell$ hypothesis, $M_{min}(K\pi\ell\nu_\ell)$, 
is required to be greater than 2.0\,GeV/$c^2$.  
Monte Carlo studies show that 
less than 5\% of observed \kstln\ decays populate 
the $M_{min}(K\pi\ell\nu_\ell)$ distribution above 2.0\,GeV/$c^2$, while 
about 70\% of \rholn\ decays populate this region
(when a pion is incorrectly assigned the mass of a kaon).
2) Information from the \v{C}erenkov counters for both
hadron candidates is used to reject about 51\% of $K\pi$ pairs,
yet keep about 92\% of $\pi\pi$ pairs.
3) Although no significant $K^{*}$ peak in $K\pi$ invariant mass 
remains after these requirements, 
the $K\pi$ mass for the hadrons is still required to be
outside the interval 0.85 to 0.93\,GeV/$c^2$.
These three cuts combined with those described earlier result in a relative
reduction factor of nearly 200 for the \kstln~mode compared with the
\rholn~mode.   

\begin{figure}[htbp]
    \epsfxsize=4.5in
    \centerline{\epsffile[72 270 522 666]{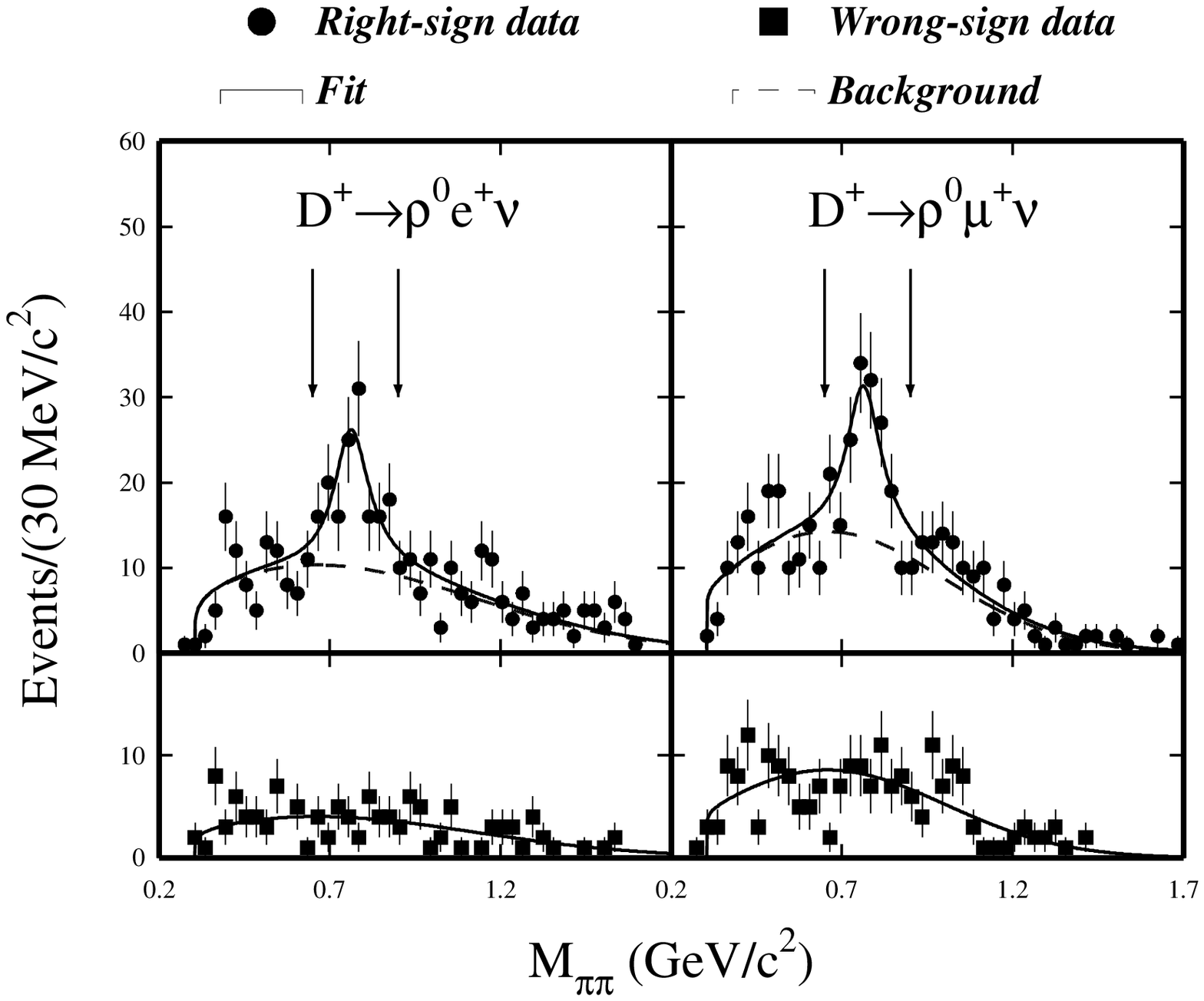}}
    \epsfverbosetrue
    \caption{$M_{\pi\pi}$ distribution for \rholn\ candidates. The
    vertical arrows indicate the mass window for the final \rholn\ 
    candidates. 
    For each leptonic mode, a simultaneous fit is made to the RS and WS data. 
    The shape for the background distribution in the
    RS data is constrained to be the same as that of WS distribution, 
    but the relative normalization is allowed to vary.
}
    \label{fig1}
\end{figure}

\begin{figure}[htbp]
    \epsfxsize=4.5in
    \centerline{\epsffile[72 270 522 666]{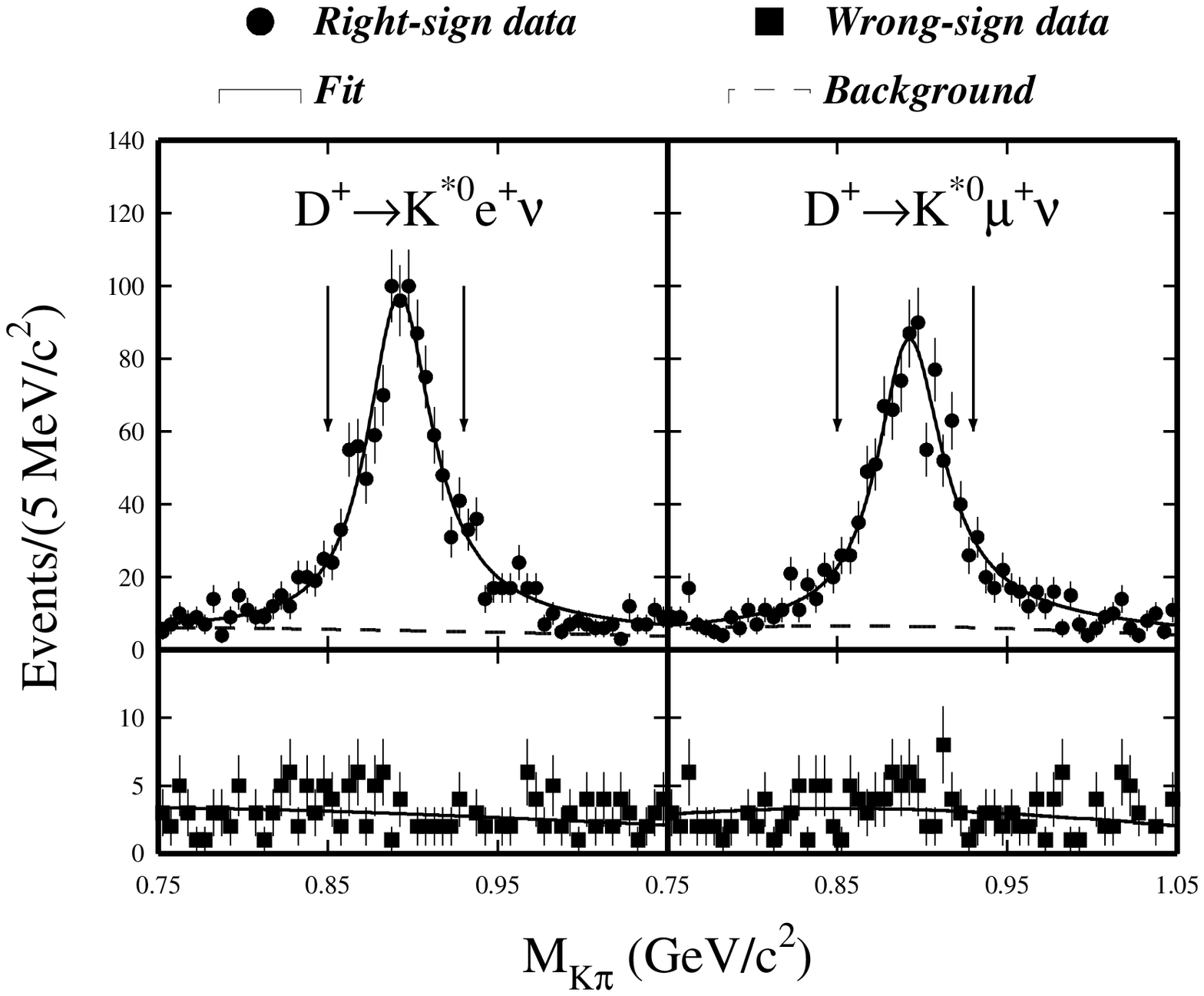}}
    \epsfverbosetrue
    \caption{$M_{K\pi}$ distribution for \kstln\ candidates. The
    vertical arrows indicate the mass window for the final \kstln\ 
    candidates.
    For each leptonic mode, a simultaneous fit is made to the RS and WS data. 
    The shape for the background distribution in the
    RS data is constrained to be the same as that of WS distribution, 
    but the relative normalization is allowed to vary.
}
    \label{fig2}
\end{figure}

Figures~\ref{fig1} and \ref{fig2} show the signals in \rholn\ and the
normalizing channel \kstln\ for both the electronic and muonic modes.
Simultaneous binned maximum likelihood fits to both the RS and
WS distributions are performed separately for the electronic and
muonic channels in both the \rholn\ and \kstln\ decays. 
Two functions are used in the fit: 
a p-wave Breit-Wigner shape describes the $\rho^0$ signal and 
the function
 $F(M) = N_0(M-m_0)^\alpha\exp{\Bigl[c_1(M-m_0)+c_2(M-m_0)^2\Bigr]}$, 
where $N_0$, $m_0$, $\alpha$, $c_1$ and $c_2$ are free parameters, 
characterizes the background under the $\rho^0$ ($K^{*0}$) peak in the
RS sample  
which is assumed to have the same distribution as the WS sample. 
The normalizations for the RS
background and the WS distribution are allowed to vary independently.
In the case of \rholn, the shape of the $\rho^0$ is modified by the
energy available in the $D^+$ decay. Thus, the $\rho^0$ mass and width
are taken from Monte Carlo simulation of the \rholn\ decay.
For the \kstln\ mode, both the $K^{*0}$
width and peak position are free parameters in the fit; the values
obtained from the fit agree with those from Monte Carlo. 
The p-wave Breit-Wigner functions from the fits are integrated 
from 0.65 to 0.90\,GeV/$c^2$ for the $\pi^+\pi^-$ invariant mass for
the \rholn\ signal,  
and from 0.85 to 0.93\,GeV/$c^2$ for the $K\pi$ invariant mass for the
\kstln\ signal. The yields for both the \rholn\ and \kstln\ channels
are listed in
Table~\ref{tab:background}.   

The efficiencies are factorized into two parts. The \v{C}erenkov particle
identification efficiencies are determined from a sample of 
 $D^+\rightarrow K^-\pi^+\pi^+$ decays from real data, where the kaon and pions
can be identified by charge alone. The rest of the 
reconstruction efficiencies and acceptances are determined from Monte
Carlo simulation. Only the relative efficiencies for \rholn\ and
\kstln\ enter our final result.
The overall efficiencies for the \rholn\ and \kstln\ channels, as well
as for the background modes, are listed in Table~\ref{tab:background}.
 
Only backgrounds which populate the $\rho^0$ region and mimic 
the $\rho^0$ resonance are troublesome in the raw \rholn\ signal. 
Using simulated hadronic charm decays from the 
channels most likely to feed into \rholn\ signal, 
we found that hadronic charm decay 
feedthrough to the \rholn\ signal is negligible.

The background contributions to the signal mainly come from real
semileptonic charm decays.  
The amount of feedthrough from them is based on efficiencies
estimated from Monte Carlo studies and 
Particle Data Group (PDG)\cite{pdg} branching
ratios unless otherwise noted. 
To estimate the backgrounds from $D_s^+$ decays, the $D_s^+$ to $D^+$ 
production cross section ratio $\sigma_{D_s^+}/\sigma_{D^+}$ is
needed. The weighted average of the measurements from hadronic charm 
production experiments\cite{prod} with a conservative error,
$0.58\pm0.15$, is used. 
The most significant semileptonic charm decay backgrounds are listed 
in Table~\ref{tab:background} along with the corresponding estimated
number of events in the \rholn\ signal.
Requiring candidates to lie between 0.65
 and 0.90\,GeV/$c^2$ in $\pi^+\pi^-$ mass effectively removes decays
 from \philn, $\phi\rightarrow K^+K^-$, as well as from \Detaln, \Fetaln. 
The contribution to the signal from these modes is negligible.
The backgrounds from non-resonant or higher-mass-resonance decays 
are negligible as well. 

After background subtraction, the final numbers of signal events are 
 $49\pm17$ for \rhoen\ and 
 $54\pm18$ for \rhomn.
The yields in the normalizing channels are
 $892\pm52$ for \ksten\ and $769\pm54$ for \kstmn.

Systematic errors associated with lepton identification are
largely cancelled in the ratio of the \rholn\ and \kstln~decay rates.
Remaining sources of systematic error are 1) uncertainties in 
the branching ratios used in background subtraction, 
2) uncertainty in the $D_s^+$ to $D^+$ production cross section ratio 
 $\sigma_{D_s^+}/\sigma_{D^+}$, 3) determination of relative efficiencies and 
4) the fitting procedure.
The uncertainties in the relative efficiencies are dominated by the
momentum dependence of the \v{C}erenkov identification and
the dependence of the \rholn\ acceptances on the form factors. 
The effects of the uncertainties in the assumptions made in the fit
are evaluated by varying 
the width of the $\rho^0$ peak and the shape of the background distribution. 
For the electronic channel, the four sources contribute approximately
equally to the
systematic error, each with a size about one third of the statistical
error. For the muonic channel, the first three sources contribute
approximately equally (about one third of the statistical error each)
while the
last source contributes an uncertainty about one sixth
that of the statistical error.   
Further studies were performed to search for other potential
systematic effects by varying the selection criteria one
at a time. No significant effects were found.

The rate for the decay $D \rightarrow V \ell^+ \nu_\ell$, where $V$ is a
vector meson, is determined in the limit of massless $\ell$ by three
form factors $A_1(q^2)$, $A_2(q^2)$ and $V(q^2)$, where $q^2$ is the
square of the four-momentum transfer from $D$ to $V$~\cite{burchat_PDG}.
The present experimental information for \kstln\ decays is usually
presented as $A_1(0)$, and the ratios of form factors 
 $r_2(0)\equiv A_2(0)/A_1(0)$ and $r_V(0)\equiv V(0)/A_1(0)$,
with an assumed $q^2$ dependence proportional to $(1-q^2/M_p^2)^{-1}$ and
 $M_p\sim 2.1$ to 2.5 GeV/$c^2$. Since we have insufficient statistics in
this experiment to measure the \rholn\ form
factors, the   simulations on which our efficiencies are based assumed
 $r_2(0)=0.82$ and $r_V(0)=2.0$, close to the present world
averages~\cite{burchat_PDG} for \kstln. We
have checked the effect on the detection efficiency of significantly
different values of the input $r_2$ and $r_V$. Specifically, assuming
 $r_2(0)=0.0$ or $r_V(0)=1.0$ changes our \rholn\ branching fraction
by less than $(10\pm 5)\%$ of itself. These variations have been
included in the systematic uncertainty.

From the background-subtracted event yields and the efficiencies for
\rholn\ and \kstln\ decays, 
the following branching ratios are determined:
\begin{eqnarray*} 
\frac{{\cal B}(\rhoen)}{{\cal B}(\ksten)} & = & (4.5\pm1.4\pm0.9)\%, \\
\frac{{\cal B}(\rhomn)}{{\cal B}(\kstmn)} & = & (5.1\pm1.5\pm0.9)\%.
\end{eqnarray*} 
We combine the results from the electronic and muonic modes, 
taking correlated errors into account, to obtain a final result of
 $$\frac{{\cal B}(\rholn\ )}{{\cal B}(\kstln )}=(4.7\pm1.3)\%,$$ 
where the error includes both statistical and systematic uncertainties.
In Table~\ref{tab:compare} our result is compared with the only 
previously-published experimental result, from E653\cite{e653-rho}, and various
theoretical predictions.  
Our result, which is sensitive mainly to the 
form factor $A_1$ at $q^2 \approx 0.5$ (GeV/$c$)$^2$, 
agrees only marginally with 
quark model predictions\cite{isgw2,jaus,wirbel}, 
but agrees well with recent 
lattice QCD calculations\cite{abada,allton,bowler,lubicz} and several
other theoretical predictions\cite{bajc,casalbuoni,ball}. 
It thus begins to discriminate among
models that are also used to predict form factors for 
 $b \rightarrow u$ semileptonic decays to extract $V_{ub}$.

We gratefully acknowledge the assistance of the staffs of Fermilab and of all
the participating institutions.  This research was supported by the Brazilian
Conselho Nacional de Desenvolvimento Cient\'{i}fico e Technol\'{o}gico,
CONACyT (Mexico), the Israeli Academy of Sciences and Humanities, 
the U.S. Department of Energy, the U.S.-Israel
Binational Science Foundation, and the U.S. National Science Foundation.
Fermilab is operated by the Universities Research Association, Inc., under
contract with the United States Department of Energy.

\begin{table}[htbp]
\caption{Numbers of 
 $D^+\rightarrow\rho^0\ell^+\nu_\ell$ and 
 $D^+\rightarrow\overline{K}^{*0}\ell^+\nu_\ell$ signal events from the fit to
the data, estimated numbers of background events, 
efficiencies ${\cal E}$ for each decay mode
and input to the calculation of backgrounds. Efficiencies include
branching ratios for non-charm decays.
}
\label{tab:background}
\begin{center}
\begin{tabular}{lcccl}
  Decay & $\ell$ & \# of events & ${\cal E}$(\%) & Input \\ \hline
 Raw $\rho^0\ell\nu_\ell$ signal & $\mu$ & $81.3\pm16.6$ &
  & Fit results, before  \\
       &$e$& $73.9\pm15.2$ & & background subtraction \\ \hline \hline
\Fetapln\ & $\mu$ & $15.1\pm6.4$ & 0.073 & CLEO
  measurement\cite{cleo:eta} for  \\ 
 $\hspace*{1.2cm}\hookrightarrow\gamma\rho,\gamma\omega$ &$e$& $14.2\pm6.0$ &
  0.064 & ${\cal B}(\Fetapln)/{\cal B}(\philn)$ \\ \hline
\kstln\  & $\mu$ & $4.1\pm1.5$ & 0.001 & PDG\cite{pdg} branching ratios\\ 
 $\hspace*{1.2cm}\leadsto\mbox{``$\pi^-$''}\pi^+$ &$e$& $3.6\pm1.1$
  & 0.001 & \\ \hline 
\philn\  & $\mu$ & $3.7\pm1.3$ & 0.017 & PDG\cite{pdg} branching ratios\\ 
 $\hspace*{1.2cm}\hookrightarrow\rho\pi,\pi^+\pi^-\pi^0$ &$e$&
  $3.5\pm1.2$& 0.014 & 
  \\ \hline
\Detapln\ & $\mu$ & $2.9\pm1.2$ & 0.088 & ISGW2\cite{isgw2}
  prediction for \\
 $\hspace*{1.2cm}\hookrightarrow\gamma\rho,\gamma\omega$ &$e$& $2.1\pm1.2$ &
  0.062 & ${\cal B}(\Detapln)/{\cal B}(\rholn)$ \\ \hline
\omegaln\ & $\mu$ & $1.2\pm0.3$ & 0.005 & ISGW2\cite{isgw2} prediction for \\
 $\hspace*{1.2cm}\hookrightarrow\pi^+\pi^-\pi^0,\pi^+\pi^-$ &$e$& $1.1\pm0.4$ &
  0.004 & ${\cal B}(\omegaln)/{\cal B}(\rholn)$ \\ \hline
\hline
\rholn\ & $\mu$ & $54\pm18$ & 0.19 & Background subtracted
  signal \\
 $\hspace*{1.2cm}\hookrightarrow\pi^+\pi^-$ &$e$& $49\pm17$ & 0.16 &
 \\ \hline
\multicolumn{5}{l}{Normalizing mode} \\ \hline
\kstln\  & $\mu$ & $769\pm54$ & 0.19 & Fit results for \\ 
 $\hspace*{1.2cm}\hookrightarrow K^-\pi^+$ &$e$
 & $892\pm52$ & 0.19 & normalizing signals\\ 
\end{tabular}
\end{center}
\end{table}

\begin{table}[htbp]
\caption{Comparison of our results with the only previously-published
experimental result and theoretical predictions. 
The PDG
{\protect\cite{pdg}} 
branching ratio for
 $D^+\rightarrow\overline{K}^{*0}\ell^+\nu_\ell$ and $D^+$ lifetime are
used to calculate the experimental decay rate for
 $D^+\rightarrow\rho^0\ell^+\nu_\ell$.  
Most theoretical results are calculated for $D^0$
decays. To compare these results for $D^0$ decay with the experimental
results for $D^+$ decay we use the relations 
 $\Gamma(D^+\rightarrow \overline{K}^{*0} \ell^+\nu_\ell) =
  \Gamma(D^0\rightarrow K^{*-} \ell^+\nu_\ell)$ and 
 $\Gamma(D^+\rightarrow \rho^0 \ell^+\nu_\ell) = 1/2\times
  \Gamma(D^0\rightarrow \rho^- \ell^+\nu_\ell)$, where the factor of 2
difference between the two decay rates arises from the $1/\sqrt{2}$
coupling of $d\bar{d}$ to the $\rho^0$.
The second column indicates the method used to obtain the results,
where QM stands for quark model, HQET for heavy quark effective
theory, SR for QCD sum rule, 
and LQCD for lattice QCD.}
\label{tab:compare}
\begin{center}
\begin{tabular}{ccccc}
Group  & Method & $\ell$  & $\frac{\Gamma(D^+\rightarrow \rho^0
                         \ell^+\nu_\ell)} 
                         {\Gamma(D^+\rightarrow K^{*0} \ell^+\nu_\ell)}$ 
 & $\Gamma(D^+\rightarrow \rho^0 \ell^+\nu_\ell)$ ($10^{10}s^{-1}$)\\
\hline
E791(this work) & Exp. & $e$    & $0.045\pm0.014\pm0.009$ 
& $0.20\pm0.07\pm0.05$ \\
E791(this work) & Exp. & $\mu$  & $0.051\pm0.015\pm0.009$ 
& $0.22\pm0.07\pm0.05$ \\
E791(this work) & Exp. & $\ell$ & $0.047\pm0.013$ & $0.21\pm0.06$ \\
E653 \cite{e653-rho} & Exp. & $\mu$ 
& $0.044^{+0.031}_{-0.025}\pm0.014$ 
& $0.19^{+0.14}_{-0.11}\pm0.07$ \\
\hline
\hline
ISGW2 \cite{isgw2} & QM & $\ell$ & 0.022 & 0.12 \\
Jaus \cite{jaus} & QM & $\ell$ & 0.030 & 0.33 \\ 
Bajc \cite{bajc} & HQET & $\ell$ & --- & $0.21\pm0.02$ \\
\hline
 & & & $\frac{1}{2}\frac{\Gamma(D^0\rightarrow \rho^- \ell^+\nu_\ell)}
                        {\Gamma(D^0\rightarrow K^{*-} \ell^+\nu_\ell)}$ 
& $\frac{1}{2}\Gamma(D^0\rightarrow \rho^- \ell^+\nu_\ell)$
($10^{10}s^{-1}$)
\\ \hline
BSW \cite{wirbel} & QM & $\ell$ & 0.037 & 0.35 \\   
ELC \cite{abada} & LQCD & $\ell$ & $0.047\pm0.032$ & $0.3\pm0.15\pm0.05$\\
APE \cite{allton} & LQCD & $\ell$ & $0.043\pm0.018$ & $0.3\pm0.1$ \\
UKQCD \cite{bowler} & LQCD & $\ell$ & 
 $0.036^{+0.010}_{-0.013}$ & $0.215\pm0.055$ \\ 
LMMS \cite{lubicz} & LQCD & $\ell$ & $0.040\pm0.011$ & $0.20\pm0.045$ \\
Casalbuoni \cite{casalbuoni} & HQET & $\ell$ & 0.06 & 0.225 \\    
Ball \cite{ball} & SR & $\ell$ & --- & $0.12\pm0.035$ \\
\end{tabular}
\end{center}
\end{table}

\end{document}